\documentclass[11pt,a4paper,oneside]{article}
\usepackage{graphicx}

\begin{document}

\title{An empirical mass formula for $\mu$ an $\tau$ leptons and
some remarks on trident production}

\author{T.\thinspace Jacobsen\\
Department of Physics, University of Oslo\\
PB 1048 Blindern\\
N-0316 Oslo, Norway}
\maketitle

\begin{abstract}
For integer values of its free parameter, an empirical formula reproduces 
fairly well the mass values of the $\mu$ and $\tau$ leptons as if they were 
excited states of the electron. Trident production might possibly be due 
to constituent collisions.
\end{abstract}

\bigskip
\begin{flushleft}
\end{flushleft}
\thispagestyle{empty}

\newpage

\section*{An empirical mass formula}
Three charged leptons are known, the electron and the heavier $\mu$ and 
$\tau$ leptons \cite{pdb}. Relations between their mass values have
been proposed \cite{bar}. It is therefore interesting to note that 
the empirical formula
\begin{center}
$m(n) = 3^{n/4} (4\pi)^{1+n} m_e$,
\end{center}
where $m_e$ is the mass of the electron, gives 
\begin{center}
$m(1) = 106.199$ MeV/c$^2$
\end{center}
which is 0.5 per cent off the mass of the $\mu$-lepton \cite{pdb}, and
\begin{center}
$m(2) = 1756.35$ MeV/c$^2$
\end{center}
which is 1.2 per cent off the mass of the $\tau$-lepton \cite{pdb}. 
That this formula reproduces the mass values of $\mu$ and of $\tau$ fairly
well for integer values of $n$ as if $n$ is a quantum number suggests that 
$\mu$ and $\tau$ are excited states of the electron. If so, the electron 
would have to be a complex system of some constituents, e.g. 
rishons/preons \cite{har,shu,dug}.

\section*{Remarks on trident production}
The electron and the proton are both stable particles. While a proton has 
constituents, has an electron constituents ? 

If so, and if they by analogy to the three quarks in a proton were two
$x$ with electric charge $-2e/3$ each, and one $y$ with electric charge 
$+e/3$, where each $x$ and $y$ has lepton number $1/3$, then a system 
with one $x$ and two $y$ could correspond to a neutrino.

An $x\bar{x}$ system with lepton number 
zero could correspond to a boson. If $x$ and $\bar{x}$ analogous to 
quarks may be ``dressed'' as proposed in Fig.\thinspace 1, an $x\bar{x}$ 
system may decay to a
$e^- e^+$ pair.

``Spectators'' and ``leading particles'' are seen in $pd$ reactions where 
a beam proton collides with the neutron in the deuteron with its proton as 
spectator \cite{bak}, and in $\bar{p}p$ reactions with a leading meson due 
to a dressed spectator quark \cite{bal}. If $e^-$ and $e^+$ have some 
structure, leading spectator constituents could by analogy be expected also 
in e.g. $e^- e^+$ collisions. If they by analogy to quarks may be 
``dressed'', they may be observed. 

If one $x$ of a colliding $e^-$ is scattered and dressed to be 
an $x\bar{x}$ boson which decays to an $e^- e^+$ pair, with the remaning 
$xy$ system as a spectator which is dressed to be an $xxy$ electron 
as proposed in Fig.\thinspace 2, the result is trident production, i.e.
\begin{center}
$e^- \rightarrow e^- e^+ e^-$.
\end{center}

Tridents have been observed \cite{fow} and discussed from theoretical point 
of view \cite{read}. Our Fig.\thinspace 2 suggests an additional production 
model. 

\section*{Summary}
Our empirical mass formula and model for trident production suggest that 
electrons may have an inner structure and some constituents.

\bigskip
\bigskip
\begin{center}
\includegraphics[width=12cm]{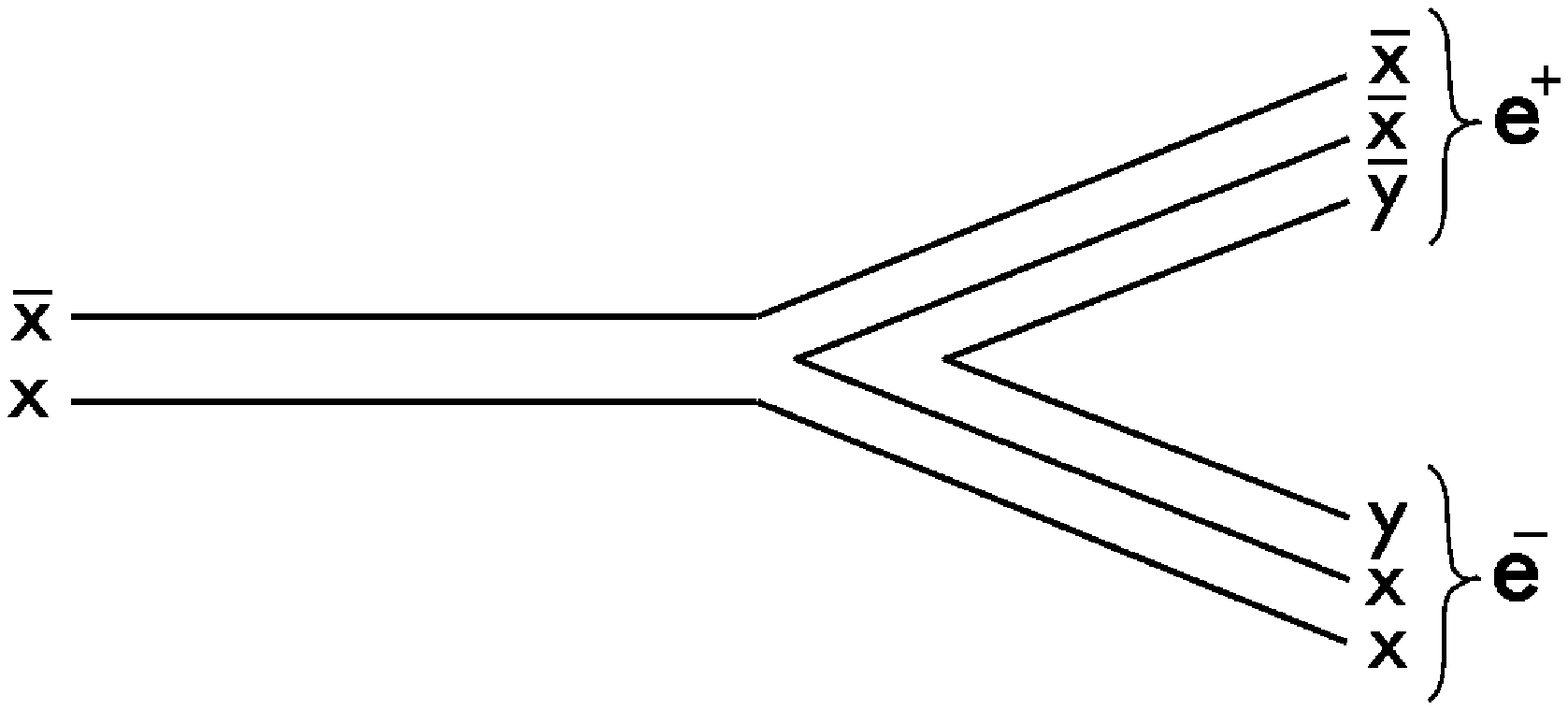}
\end{center}
\begin{center}
{\bf Fig.\thinspace 1.}\\ 

Dressing of $x \bar{x}$ to an $e^- e^+$ pair as 
proposed in the text.
\end{center}

\bigskip
\begin{center}
\includegraphics[width=12cm]{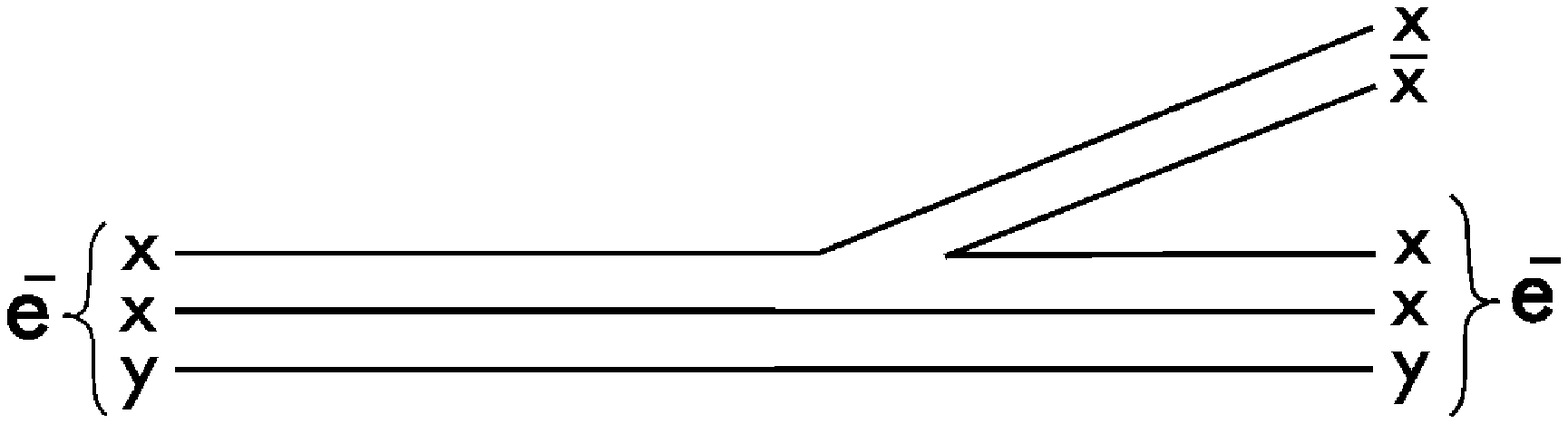}
\end{center}
\begin{center}
{\bf Fig.\thinspace 2.}\\ 

Separation and dressing of the assumed constituents 
of an electron to an electron and an $x \bar{x}$ pair as proposed in the text. 
\end{center}


\newpage
\bigskip
\begin{thebibliography}{99}

\bibitem{pdb}
     Particle Data Booklet 2004, LBNL and CERN, and 
     Phys. Lett. B {\bf 592}, 1(2004).
\bibitem{bar}
     A.O.\thinspace Barut, Phys. Rev. Lett. {\bf 42}, 1251 (1979).
     P.\thinspace Caldirola, Lett. Nuovo Cim. {\bf 27}, 225 (1980).
     H.\thinspace Terazawa and M.\thinspace Yasue, 
     Phys. Lett. B {\bf 307}, 383 (1993).
     Y.\thinspace Koide, arXiv:hep-ph/0506247 v1, 2005.
\bibitem{har}
     H.\thinspace Harari, Phys. Lett. B {\bf 86}, 83 (1979), and Scientific
     American {\bf 248}, 48 (1983).
\bibitem{shu}
     M.A.\thinspace Shupe, Phys. Lett. B {\bf 86}, 87 (1979).
\bibitem{dug}
     J.-J.\thinspace Dugne, S.\thinspace Fredriksson, 
     and J.\thinspace Hansson, Europhys. Lett. {\bf 57}, 188 (2002).
     S.\thinspace Fredriksson, arXiv:hep-ph/0309213 v2, Sep. 2003.
     A.J.\thinspace Buchmann and M.L.\thinspace Schmid, 
     Phys. Rev. D {\bf 71}, 055002(2005), and references therein.
     A.\thinspace Breakstone, e-Print Arhive : physics/0602118, 2006,
     and references therein.
\bibitem{bak}
     V.\thinspace Bakken and T.\thinspace Jacobsen, 
     Nuovo Cim. A {\bf 61}, 219 (1981), and references therein.
\bibitem{bal}
     F.\thinspace Balestra et al., Phys. Lett. B {\bf 217}, 43 (1989).
     F.\thinspace Balestra et al., Phys. Scripta {\bf 43}, 9 (1991).
\bibitem{fow}
     P.H.\thinspace Fowler, D.H.\thinspace Perkins and 
     C.F.\thinspace  Powell, The Study of Elementary Particles by the 
     Photographic Method, Pergamon Press, London 1959, and references therein.
\bibitem{read}
     G.\thinspace Reading Henry, Phys. Rev. {\bf 154}, 1534 (1967),
     K.A.\thinspace Thompson and P.\thinspace Chen, SLAC PUB 7776 (1998),
     T.\thinspace Adams et al., hep-ex/9811012, 9 Nov 1998, and references
     therein.

\end{thebibliography}
\end{document}